\documentclass[ reprint, amsmath,amssymb, prl,citeautoscript]{revtex4-1}

\begin{document}
\title{Linear representations of $SU(2)$ described by using Kravchuk polynomials}
\author{Nicolae Cotfas}
\affiliation{University of Bucharest,  Faculty of Physics,\\ P.O. Box MG-11, 077125 Bucharest, Romania}
\email{ncotfas@yahoo.com}
\homepage{http://fpcm5.fizica.unibuc.ro/~ncotfas/}
\begin{abstract}
We show that a new unitary transform with characteristics almost similar to those of the finite Fourier transform can be defined in any finite-dimensional Hilbert space. 
It is defined by using the Kravchuk polynomials, and we call it Kravchuk transform. 
Some of its properties are investigated and used  in order to obtain a simple alternative description for the irreducible  representations of the Lie algebra $su(2)$ and group $SU(2)$.  
Our approach offers a deeper insight into the structure of the linear representations of $SU(2)$ and new possibilities of computation, very useful in applications in quantum mechanics, quantum information, signal and image processing.
 
\end{abstract}
\maketitle

\section{Introduction}
The Hilbert space $\mathbb{C}^d$, of dimension $d\!=\!2j\!+\!1$, can be regarded as the space of all the functions of the form
\[
\psi\!:\!\{-j, -j\!+\!1,\, ...\, ,j\!-\!1, j\}\!\longrightarrow \!\mathbb{C},
\]
considered with the scalar product defined as
\begin{equation}\label{scalprod}
\begin{array}{l}
\langle \varphi ,\psi \rangle =\sum\limits_{n=-j}^j \overline{\varphi(n)}\, \psi(n).
\end{array}
\end{equation}
The finite Fourier transform 
$F\!:\!\mathbb{C}^d\!\longrightarrow \!\mathbb{C}^d:\psi \mapsto F[\psi ]$,
\begin{equation}
\begin{array}{l}
F[\psi](k)\!=\!\frac{1}{\sqrt{d}}\sum\limits_{n=-j}^j\!\! {\rm e}^{-\frac{2\pi {\rm i}}{d}kn}\,\psi(n),
\end{array}
\end{equation}
plays a fundamental role in quantum mechanics, signal and image processing.
Its inverse is the adjoint transform
\begin{equation}
\begin{array}{l}
F^{+}[\psi](k)\!=\!\frac{1}{\sqrt{d}}\sum\limits_{n=-j}^j\!\! {\rm e}^{\frac{2\pi {\rm i}}{d}kn}\,\psi(n),
\end{array}
\end{equation}
and $F^{4}=\mathbb{I}$, 
where $\mathbb{I}$ is the identity operator   $\mathbb{I}\psi  =\psi  .$ In the case of a quantum system with Hilbert space $\mathbb{C}^d$,
the self-adjoint operator 
$Q\!:\!\mathbb{C}^d\!\longrightarrow \!\mathbb{C}^d:\psi \mapsto Q\psi $,
\begin{equation}
\begin{array}{l}
(Q\psi )(n)\!=\!n\, \psi(n),
\end{array}
\end{equation} 
is usually regarded as a coordinate operator and 
\begin{equation}
P\!=\!F^+QF
\end{equation}
as a momentum operator  \cite{CD,CGV,Schwinger,Vourdas}. 
 
We show that a new remarkable unitary transform
\begin{equation}
K\!:\!\mathbb{C}^d\!\longrightarrow \!\mathbb{C}^d
\end{equation}  
can be defined  by using the Kravchuk polynomials. Our transform satisfies the unexpected relation $K^{3}=\mathbb{I}$, and     
\[
-{\rm i}\, Q,\qquad -{\rm i}\, K^+QK,\qquad -{\rm i}\, KQK^+
\] 
form a basis of a Lie algebra isomorphic to $su(2)$.
This allows us to obtain a new description for the linear representations of $SU(2)$ and $SO(3)$. 
In terms of Kravchuk functions, the matrix elements of the irreducible representations  have simpler mathematical expressions.

In conclusion, we define a new unitary transform K, similar to the finite Fourier transform F. We have $F^4=\mathbb{I}$, respectively $K^3=\mathbb{I}$, and in both cases, the definitions of the direct and inverse transforms are almost identical.
The new transform $K$  allows a new description of the linear representations of su(2), SU(2), SO(3),
and new developments in all the mathematical models based on these representations.
New models in physics, quantum information, quantum finance \cite{LC}, signal and image processing  can be obtained by using $K$ instead of  $F$.

\section{Kravchuk polynomials}

The functions 
\[
K_{-j},\, K_{-j+1}, ... , K_{j-1}, K_{j}\!:\!\{-j,-j\!+\!1,...,j\!-\!1,j\}\longrightarrow \mathbb{R}
\]
satisfying the polynomial relation \cite{Vilenkin}
\begin{equation}\label{KPgen}
(1\!-\!X)^{j+k}(1\!+\!X)^{j-k}=\sum_{m=-j}^{j}K_m(k)\, X^{j+m}
\end{equation}
are called {\em Kravchuk polynomials}.
The first three of them are: $K_{-j}(k)\!=\!1$, \ $K_{-j+1}(k)\!=\!-2k$, \ 
$K_{-j+2}(k)\!=\!2k^2\!-\!j$. 

By admitting that
\[
\begin{array}{l}
\frac{1}{\Gamma (n)}=0 \qquad \mbox{for}\quad n\!\in \!\{ 0,-1,-2,...\}
\end{array}
\]
 and using the binomial coefficients
\begin{equation}
\begin{array}{rl}
C_m^n & \!\!\!\!=\!\frac{\Gamma (m\!+\!1)}{\Gamma (n\!+\!1)\, \Gamma (m\!-\!n\!+\!1)}\\[3mm]
 & \!\!\!\!=\!
\left\{ 
\begin{array}{cll}
\frac{m!}{n!\, (m\!-\!n)!} & \mbox{for} & n\!\in \!\{ 0,1,2, ...,m\},\\[2mm]
0 & \mbox{for} & n\!\in \!\mathbb{Z}\backslash\{ 0,1,2, ...,m\}.
\end{array}\right.
\end{array}
\end{equation}
the Kravchuk polynomials can be defined as
\begin{equation}\label{defKm}
K_m(k)=\sum_{n=0}^{j+m}(-1)^n\, C_{j+k}^n\, C_{j-k}^{j+m-n}.
\end{equation}

The hypergeometric function 
\begin{equation}\label{hipge}
{}_2F_1\left(\left.
\begin{array}{c}
a,\, b\\
c
\end{array}\right|z\right)=\sum_{k=0}^\infty \frac{(a)_k\, (b)_k}{(c)_k}\frac{z^k}{k!}, 
\end{equation}
where 
\[
(\alpha)_k\!=\!\alpha(\alpha\!+\!1)...(\alpha\!+\!k\!-\!1)\!=\!\frac{\Gamma(\alpha\!+\!k)}{\Gamma(\alpha)},
\]
satisfies the relation \cite{Nikiforov}
\begin{equation}
\begin{array}{rl}
{}_2F_1\left(\left.\!\!
\begin{array}{c}
-n,\, \beta\\
\gamma
\end{array}\right|z\right) & \!\!\!=\frac{\Gamma (\gamma )\, \Gamma(\gamma \!-\!\beta \!+\!n)}{\Gamma (\gamma \!+\!n)\, \Gamma(\gamma\!-\!\beta)}\\[2mm]
 &  \times {}_2F_1\left(\left.\!\!
\begin{array}{c}
-n,\, \beta\\
\beta\!-\!\gamma\!-\!n\!+\!1
\end{array}\right|1\!-\!z\right).
\end{array}
\end{equation}
Since $(-\alpha )_k\!=\!(-1)^k\, \Gamma(\alpha \!+\!1)/\Gamma(\alpha\!-\!k\!+\!1)$, the relation
\[ 
\begin{array}{rl}
\mbox{}\qquad 
{}_2F_1\left(\left.\!\!\!\!
\begin{array}{c}
-j\!-\!m,\, -j\!-\!k\\
1\!-\!k\!-\!m
\end{array}\!\!\right|\!-1\right)& \!\!\!=\frac{\Gamma (1\!-\!k\!-\!m )\, \Gamma(2j\!+\!1)}{\Gamma (j\!-\!k\!+\!1)\, \Gamma(j\!-\!m\!+\!1)}\\[2mm]
 &  \times{}_2F_1\left(\left.\!\!\!\!
\begin{array}{c}
-j\!-\!m,\, -j\!-\!k\\
-2j
\end{array}\!\!\right|\!2\right)
\end{array}
\]
can be written as
\begin{equation}\label{hypKrav}
\begin{array}{l}
K_m(k)\!=\!C_{2j}^{j+m}\,
{}_2F_1\!\!\left(\left.\!\!\!\!
\begin{array}{c}
-j\!-\!m,\, -j\!-\!k\\
-2j
\end{array}\!\!\right|\!2\right).
\end{array}
\end{equation}

From the polynomial relation
\[
\begin{array}{l}
\sum\limits_{m,n=-j}^j\left(\frac{1}{2^{2j}}\sum\limits_{k=-j}^j C_{2j}^{j+k}\, K_m(k)\, K_n (k)\right)X^{j+m}\, Y^{j+n}\\[2mm]
=\frac{1}{2^{2j}}\sum\limits_{k=-j}^j C_{2j}^{j+k}\, \sum\limits_{m=-j}^{j}K_m(k)\, X^{j+m}\, \sum\limits_{n=-j}^{j}K_n (k)\, Y^{j+n }\\[2mm]
=\!\frac{1}{2^{2j}}\!\!\sum\limits_{k=-j}^j\!\! C_{2j}^{j+k}\,(1\!-\!X)^{j+k}(1\!+\!X)^{j-k}(1\!-\!Y)^{j+k}(1\!+\!Y)^{j-k} \\[2mm]
=\frac{1}{2^{2j}}[(1\!-\!X)(1\!-\!Y)+(1\!+\!X)(1\!+\!Y)]^{2j}= (1+XY)^{2j}
\\[2mm]
=\sum\limits_{m=-j}^j C_{2j}^{j+m}X^{j+m} Y^{j+m}.
\end{array}
\]
it follows the well-known relation \cite{Nikiforov,Vilenkin}
\begin{equation}
\begin{array}{l}
\frac{1}{2^{2j}}\sum\limits_{k=-j}^j C_{2j}^{j+k}\, K_m(k)\, K_n (k)= C_{2j}^{j+m}\, \delta_{mn}.
\end{array}
\end{equation}
We extend  the set $\{K_m\}_{m=-j}^j$ by admitting that
\[
K_m=0\quad \text{for}\quad m\!\in \!\mathbb{Z}\backslash\{-j,-j\!+\!1,...,j\!-\!1,j\}.
\]
With this convention, by differentiating (\ref{KPgen}) we get 
\begin{equation}\label{KPrec}
\begin{array}{l}
(j\!+\!m\!+\!1)\, K_{m+1}(k)\\[2mm]
\qquad+(j\!-\!m\!+\!1)\, K_{m-1}(k)\!=\!-2k\, K_{m}(k).
\end{array}
\end{equation}
For the half-integer powers we use the definiton  
\begin{equation} 
\qquad \qquad z^k=|z|^k \, {\rm e}^{{\rm i}k\, {\rm arg}(z)}.
\end{equation}
{\bf Theorem 1}. {\em Kravchuk polynomials satisfy the relation
\begin{equation}\label{new-rel-p} 
\sum_{k=-j}^j (-{\rm i})^k\, K_m(k)\, K_k(n)=2^j\, {\rm i}^{j+m}\, {\rm i}^{j+n}\,K_m(n)
\end{equation}
for any   $m,n \!\in \!\{-j,-j\!+\!1,...,j\!-\!1,j\}$.}\\[5mm]
{\em Proof}. Direct consequence of the polynomial relation
\[
\begin{array}{l}
\sum\limits_{m=-j}^{j}\left(\sum\limits_{k=-j}^{j}(-{\rm i})^{j+k}K_m(k)\, K_k(n)\right) X^{j+m}\\[3mm]
\qquad =\sum\limits_{k=-j}^{j}(-{\rm i})^{j+k}(1\!-\!X)^{j+k}(1\!+\!X)^{j-k}\, K_k(n)\\[3mm]
\qquad =(1\!+\!X)^{2j}\sum\limits_{k=-j}^{j}K_k(n)\left(\frac{X-1}{X+1}{\rm i} \right)^{j+k}.\\[3mm]
\qquad =(1\!+\!X)^{2j}\left(1-\frac{X-1}{X+1}{\rm i}\right)^{j+n}
\left(1+\frac{X-1}{X+1}{\rm i}\right)^{j-n}\\[3mm]
\qquad =\left(1\!+\!{\rm i}+(1\!-\!{\rm i})X \right)^{j+n}
\left(1\!-\!{\rm i}+(1\!+\!{\rm i})X \right)^{j-n}\\[3mm]
\qquad=(1\!+\!{\rm i})^{j+n}(1\!-\!{\rm i})^{j-n}
(1\!-\!{\rm i}X)^{j+n}(1\!+\!{\rm i}X)^{j-n}\\[3mm]
\qquad =2^j (-{\rm i})^{j}\, {\rm i}^{j+n}\, \sum\limits_{m=-j}^{j}K_m(n)({\rm i}X)^{j+m}.
\qquad \Box
\end{array}
\]
By using (\ref{hypKrav}), the relation (\ref{new-rel-p}) can be written as
\begin{equation}
\begin{array}{r}
\sum\limits_{k=-j}^j(-{\rm i})^k \, \frac{(2j)!}{(j\!+\!k)!\, (j\!-\!k)!}\ 
{}_2F_1\left(\left.\!\!
\begin{array}{c}
-j\!-\!m,\, -j\!-\!k\\
-2j
\end{array}\right|2\right) \\[5mm]
\qquad \times {}_2F_1\left(\left.\!\!
\begin{array}{c}
-j\!-\!k,\, -j\!-\!n\\
-2j
\end{array}\right|2\right)\\[5mm]
=2^j\, {\rm i}^{j+m}\, {\rm i}^{j+n}\,
{}_2F_1\left(\left.\!\!
\begin{array}{c}
-j\!-\!m,\, -j\!-\!n\\
-2j
\end{array}\right|2\right),
\end{array}
\end{equation}
and is a special case for  (12) from \cite{E} and  (5.5) from \cite{J}.

\section{Kravchuk functions}

The space $\mathbb{C}^d$ can be regarded as a subspace of the space of all the functions of the form $\psi:\mathbb{Z}\longrightarrow \mathbb{C}$ by identifying it either with the space 
\begin{equation}
\ell ^2(\mathbb{Z}_d)\!=\!\{ \ \psi \!:\!\mathbb{Z}\!\longrightarrow \!\mathbb{C}\ |\ \psi(n\!+\!d)\!=\!\psi(n) \ \mbox{for any} \ n\!\in \!\mathbb{Z}\ \}
\end{equation}
of periodic functions of period $d$ or with the space 
\begin{equation}
\ell ^2[-j,j]\!=\!\{ \, \psi \!:\!\mathbb{Z}\!\longrightarrow \!\mathbb{C}\ \, |\ \, \psi(n)\!=\!0\ \ \mbox{for}\ \ |n|>j\, \}
\end{equation}
of all the functions null outside $\{-j, -j\!+\!1,\, ...\, ,j\!-\!1, j\}$.
The use of $\ell ^2(\mathbb{Z}_d)$ or $\ell ^2[-j,j]$ allows us to define new mathematical objects and to obtain new results.
The functions $\mathfrak{K}_{m}\!:\!\mathbb{Z}\longrightarrow \mathbb{R}$, 
defined as
\begin{equation}\label{KFdef}
\mathfrak{K}_m(k)=
\frac{1}{2^j}\sqrt{\frac{C_{2j}^{j+k}}{C_{2j}^{j+m}}}\,K_m(k) 
\end{equation}
can be expressed in terms of the hypergeometric function
\begin{equation}
\mathfrak{K}_m(k)=\frac{1}{2^j}\sqrt{C_{2j}^{j+m}\, C_{2j}^{j+k}}\,\, {}_2F_1\!\!\left(\left.\!\!\!\!
\begin{array}{c}
-j\!-\!m,\, -j\!-\!k\\
-2j
\end{array}\!\!\right|\!2\right),
\end{equation}
belong to $\ell ^2[-j,j]$, and satisfy the relations 
\[
\mathfrak{K}_m(n)\!=\!\mathfrak{K}_n(m),\qquad 
\mathfrak{K}_m(-n)\!=\!(-1)^{j+m}\mathfrak{K}_m(n).
\]
The functions $\{\mathfrak{K}_{m}\}_{m=-j}^j$, called {\em Kravchuk functions}, form an orthonormal basis in $\ell ^2[-j,j]$, that is, we have
\begin{equation}
 \langle \mathfrak{K}_m|\mathfrak{K}_n \rangle =\delta_{mn},\qquad \sum_{m=-j}^j|\mathfrak{K}_m\rangle \langle \mathfrak{K}_m|=\mathbb{I}.
  \end{equation}
  
  From (\ref{KPrec}) we get the relation
\begin{equation}\label{KFrec}
\begin{array}{l}
\sqrt{(j\!-\!m)(j\!+\!m\!+\!1)}\ \mathfrak{K}_{m+1}(k)\\[3mm]
\qquad +\!\sqrt{(j\!+\!m)(j\!-\!m\!+\!1)}\ \mathfrak{K}_{m-1}(k)\!=\!-2k\, \mathfrak{K}_m(k)
\end{array}
\end{equation}
and its direct consequence
\begin{equation}\label{ortogKF}
-2\!\!\sum\limits_{k=-j}^j  \!\!k\, \mathfrak{K}_m(k)\, \mathfrak{K}_n (k)\!=\!\left\{\!\!\!
\begin{array}{lll}
\sqrt{(j\!+\!m)(j\!-\!m\!+\!1)} & \mbox{for} & n\!=\!m\!-\!1\\[2mm]
\sqrt{(j\!-\!m)(j\!+\!m\!+\!1)}& \mbox{for} & n\!=\!m\!+\!1\\[2mm]
 \ 0 &  \mbox{for} & n\!\not=\!m\pm 1.
  \end{array}\right.
\end{equation}  
The relation (\ref{KFrec}) can also be written in the form \cite{AAW,AW,Vilenkin}
\begin{equation}\label{recK2}
\begin{array}{l}
\sqrt{(j\!-\!m)(j\!+\!m\!+\!1)}\mathfrak{K}_{k}(m\!+\!1)\\[3mm]
\quad +\sqrt{(j\!+\!m)(j\!-\!m\!+\!1)}\mathfrak{K}_{k}(m\!-\!1)\!=\!-2k\, \mathfrak{K}_k(m).
\end{array}
\end{equation} 
{\bf Theorem 2}. {\em The Kravchuk functions satisfy the relation
\begin{equation}\label{new-rel-f} 
\sum\limits_{k=-j}^j (-{\rm i})^{k}\, \mathfrak{K}_{m}(k)\, \mathfrak{K}_{k}(n)={\rm i}^{j+m}\, {\rm i}^{j+n}\, \mathfrak{K}_{m}(n) 
\end{equation}
 for any   $m,n \!\in \!\{-j,-j\!+\!1,...,j\!-\!1,j\}$.}\\[5mm]
{\em Proof}. Direct consequence of (\ref{new-rel-p}) and (\ref{KFdef}).\qquad $\Box$

\section{Kravchuk transform}

The functions $\{ \delta_m\}_{m=-j}^j$, defined by the relation 
\begin{equation}\label{defbasis}
\delta_m\!:\!\mathbb{Z}\!\longrightarrow \!\mathbb{C},\qquad 
\delta_m (k)\!=\!\delta_{km}\!=\!\left\{ 
\begin{array}{ccc}
1 & \mbox{for} & k\!=\!m,\\[1mm]
0 & \mbox{for} & k\!\neq \!m,
\end{array} \right.
\end{equation}
form an orthonormal basis in the Hilbert space $\mathbb{C}^d$.  
With the traditional notation $|j;m\rangle $ instead of $\delta_m$,    we have  
 \begin{equation}
  \begin{array}{l}
  \langle j;m|j;n \rangle \!=\!\delta_{mn}\quad \mbox{and} \quad
 \sum\limits_{m=-j}^j|j;m\rangle \langle j;m|\!=\!\mathbb{I}.
 \end{array}
  \end{equation}
The transform $K:\ell ^2[-j,j]\longrightarrow \ell ^2[-j,j]$, 
\begin{equation}\label{KT}
K=(-1)^{2j}\sum_{n,m=-j}^j {\rm i}^{n}\, \mathfrak{K}_{-n}(m)\ |j;n\rangle\langle j;m|, 
\end{equation}
we call {\em Kravchuk transform}, is a unitary operator.\\
The direct and inverse transforms are quite identical:
\[
\begin{array}{l}
K[\psi ](n)\, =\, (-1)^{2j}\sum\limits_{m=-j}^j \ {\rm i}^{n}\,(-1)^{j+m}\  \mathfrak{K}_{n}(m)\,\psi(m)\\[2mm]
K^+[\psi ](n)\!=\!(-1)^{2j}\!\!\sum\limits_{m=-j}^j (-{\rm i})^{m}\,(-1)^{j+n} \mathfrak{K}_{n}(m)\,\psi(m).
\end{array}
\]
The operator $Q\!:\!\ell ^2[-j,j]\rightarrow \ell ^2[-j,j]$, $(Q\psi )(n)\!=\!n\, \psi(n)$,
admitting the spectral decomposition
\begin{equation}
\begin{array}{l}
Q=\sum\limits_{n=-j}^j n\ |j;n\rangle\langle j;n|,
\end{array}
\end{equation}
can be regarded as a coordinate operator in $\ell ^2[-j,j]$.\\[1mm]
{\bf Theorem 3}. {\em Kravchuk transform satisfies the relations}
\begin{equation}
K^3=\mathbb{I}
\end{equation}
\begin{equation}\label{JxJyJz}
K^+QK^+QK^+-KQKQK={\rm i}\, Q.
\end{equation}
{\em Proof}. A consequence of the equality (\ref{new-rel-f}) is the relation
\[
\begin{array}{l}
K^2 =\sum\limits_{n,m=-j}^j\sum\limits_{k=-j}^j {\rm i}^{m+k}\, \mathfrak{K}_{-m}(k)\, \mathfrak{K}_{-k}(n)\, |j;m\rangle \langle j;n|\\[2mm]
=\!\!\!\!\sum\limits_{n,m=-j}^j\!\!\!\!\!(-1)^{j+n}\, {\rm i}^{m-j} (-{\rm i})^j\!\!\!\sum\limits_{k=-j}^j\!\!\! (-{\rm i})^{k} \mathfrak{K}_{m}(k)\, \mathfrak{K}_{k}(n)\, |j;m\rangle \langle j;n|\\[2mm]
=\sum\limits_{n,m=-j}^j(-1)^{j+m+n}\, {\rm i}^{j+n} (-{\rm i})^j\, \mathfrak{K}_{m}(n)\, |j;m\rangle \langle j;n|\\[2mm]
=\sum\limits_{n,m=-j}^j(-1)^{n}\, {\rm i}^{j+n} (-{\rm i})^j\, \mathfrak{K}_{-n}(m)\, |j;m\rangle \langle j;n|=K^+
\end{array}
\]
which shows that $K^3=\mathbb{I}$. From (\ref{ortogKF}) and 
\[
\begin{array}{l}
QK^+QK|j;m\rangle=\sum\limits_{n=-j}^j (-1)^{m+n}n\\
  \qquad \qquad \qquad \qquad \qquad   \times \!\!\sum\limits_{k=-j}^j  k\, \mathfrak{K}_m(k)\, \mathfrak{K}_n (k)\, |j;n\rangle \\
K^+QKQ|j;m\rangle=  \sum\limits_{n=-j}^j (-1)^{m+n}m\\
 \qquad \qquad \qquad \qquad \qquad   \times \!\!\sum\limits_{k=-j}^j  k\, \mathfrak{K}_m(k)\, \mathfrak{K}_n (k)\, |j;n\rangle \\
KQK^+|j;m\rangle=  \sum\limits_{n=-j}^j(- {\rm i})^m\, {\rm i}^{n}\\
 \qquad \qquad \qquad \qquad \qquad  \times \!\!\sum\limits_{k=-j}^j  k\, \mathfrak{K}_m(k)\, \mathfrak{K}_n (k)\, |j;n\rangle
\end{array}
\]
it follows the relation
\[
QK^+QK-K^+QKQ={\rm i}\, KQK^+
\]
equivalent to (\ref{JxJyJz}).\qquad $\Box$

\section{The irreducible representations of the Lie algebra $su(2)$ and group $SU(2)$}
%
%
{\bf Theorem 4}. {\em The self-adjoint operators
\begin{equation}\label{Jx-Jy-Jz}
J_z=Q,\qquad J_x=K^+QK,\qquad J_y=KQK^+
\end{equation}
satisfy the relations 
\begin{equation}\label{Jxyz}
[J_x,J_y]={\rm i}J_z,\quad [J_y,J_z]={\rm i}J_x,\quad [J_z,J_x]={\rm i}J_y
\end{equation}
and define a linear representation of $su(2)$ in $\ell ^2[-j,j]$.}\\[3mm]
{\em Proof}.  Each of the relations (\ref{Jxyz}) is equivalent to (\ref{JxJyJz}).\   $\Box$\\[3mm]
The operators $J_{\pm}=J_x\pm {\rm i}\, J_y$ verify the relations
\begin{equation}
[J_z,J_{\pm}]=\pm J_{\pm},\qquad [J_-,J_+]=-2J_z,
\end{equation}
and, by using (\ref{ortogKF}), one can prove  that
\begin{equation}
\begin{array}{l}
J_z|j;m\rangle =m\, |j;m\rangle \\[1mm]
J_+|j;m\rangle =\sqrt{(j\!-\!m)(j\!+\!m\!+\!1)}\ |j;m\!+\!1\rangle \\[1mm]
J_-|j;m\rangle =\sqrt{(j\!+\!m)(j\!-\!m\!+\!1)}\ |j;m\!-\!1\rangle .
\end{array}
\end{equation}
In certain cases, it is useful to describe the structure of $J_z$, $J_x$, $J_y$ by using only two operators. Beside 
\[
J_z,\quad J_x\!=\!\frac{1}{2}(J_{+}+J_{+}^{+}),\quad J_y\!=\!\frac{1}{2{\rm i}}(J_{+}-J_{+}^{+})
\]
we have now the alternative representation (\ref{Jx-Jy-Jz}),
more advantageous when we pass form $su(2)$ to  $SU(2)$ and $SO(3)$.
The operator $J_x$ admits the  decomposition
\[
J_x\!=\!K^+\!\!\!\sum_{k=-j}^j \!\! k\,|j;k\rangle \langle j;k|K=\!\!\!\sum_{k=-j}^j\!\! k\, |\mathfrak{K}_{-k}\rangle \langle \mathfrak{K}_{-k}|.
\]
and consequently
\[
\begin{array}{l}
{\rm e}^{-{\rm i}\beta J_x}=\!\sum\limits_{k=-j}^j {\rm e}^{-{\rm i}\beta k}\, |\mathfrak{K}_{-k}\rangle \langle \mathfrak{K}_{-k}|=\!\sum\limits_{k=-j}^j {\rm e}^{{\rm i}\beta k}\, |\mathfrak{K}_{k}\rangle \langle \mathfrak{K}_{k}|\\[2mm]
\qquad \quad =\sum\limits_{m,n=-j}^j\sum\limits_{k=-j}^j{\rm e}^{{\rm i}\beta k}\,\mathfrak{K}_{k}(m)\, \mathfrak{K}_{k}(n)\,|j;m\rangle \langle j;n|.
\end{array}
\]
In the case of the representation of $SU(2)$ in $\ell^2[-j,j]$, 
the element with Euler angles $\alpha $, $\beta $, $\gamma $ corresponds to
\[
\begin{array}{l}
{\rm e}^{-{\rm i}\alpha J_z}{\rm e}^{-{\rm i}\beta J_x}{\rm e}^{-{\rm i}\gamma J_z}\!=\sum\limits_{m,n=-j}^j {\rm e}^{-{\rm i}(\alpha m+\gamma n)}\\[2mm]
 \qquad \qquad  \qquad \qquad \quad \times \!\!\sum\limits_{k=-j}^j\!\!{\rm e}^{{\rm i}\beta k}\,\mathfrak{K}_{m}(k)\, \mathfrak{K}_{n}(k)\,|j;m\rangle \langle j;n|.
\end{array}
\]
We think that, in certain applications, our description of the linear representations of $su(2)$ and $SU(2)$ is more advantageous  than other known descriptions \cite{M,Nikiforov,T,Vilenkin}. 
The {\em spin coherent states} \cite{Gazeau,Perelomov}, that is, the orbit of $SU(2)$ passing through $|j;-j\rangle $, is formed by the states
\[
\begin{array}{l}
|\alpha, \beta \rangle ={\rm e}^{-{\rm i}\alpha J_z}{\rm e}^{-{\rm i}\beta J_x}{\rm e}^{-{\rm i}\gamma J_z}|j;-j\rangle \\[2mm]
\qquad \quad =\!\frac{{\rm e}^{{\rm i}j\gamma }}{2^j}\sum\limits_{m=-j}^j\!\! {\rm e}^{-{\rm i}\alpha m}\!\!\sum\limits_{k=-j}^j\!\!{\rm e}^{{\rm i}\beta k}\,\sqrt{C_{2j}^{j+k}}\ \mathfrak{K}_{m}(k)\,|j;m\rangle .
\end{array}
\]
Evidently, we can neglect the unimodular factor ${\rm e}^{{\rm i}j\gamma }$.

\section{Quantum systems with finite dimensional Hilbert space}

In the continuous case, the coordinate operator $(\hat{q} \psi)(q)\!=\!q\, \psi(q)$ and the momentum operator $\hat{p}\!=\!-{\rm i}\, \frac{{\rm d}}{{\rm d}q}$ satisfy the relation $\hat{p}\!=\!\mathcal{F}^+\hat{q}\mathcal{F}$, where 
\begin{equation}
\mathcal{F}[\psi](p)=\frac{1}{\sqrt{2\pi }}\int_{-\infty}^{\infty}{\rm e}^{-{\rm i}pq}\, \psi(q)
\end{equation}
is the Fourier transform. In the finite-dimensional case,  $P=F^+QF$
defined by using the finite Fourier transform,
is usually regarded as a momentum operator \cite{CGV,Schwinger,Vourdas}.

By following Atakishyiev, Wolf {\em et al.} \cite{AAW,AW,WK}, we can consider
$\tilde{P}=K^+QK$ as a second candidate for the role of momentum operator.
In this case $[Q,\tilde{P}]=iJ_y$, and
\begin{equation}
\tilde{H}\!=\!\frac{1}{2}\tilde{P}^2+\frac{1}{2}Q^2=\frac{j(j\!+\!1)}{2}-\frac{1}{2}J_y^2
\end{equation}
corresponds to the quantum harmonic oscillator.\\ 
The function $K|j;m\rangle$ is an eigenstate of $\tilde{H}$ for any $m$.

The author is grateful to J. Van der Jeugt for letting him know that the relation (\ref{new-rel-p}) from Theorem 1 is a special case of a known identity concerning the hypergeometric function.

\newpage
\begin{widetext}
\section*{Supplementary Information}
\begin{itemize}
\item
\noindent The reader can directly check the relation (\ref{new-rel-p})  by using  the program in Mathematica 
{\small
\begin{verbatim}
j=10   
K[m_,n_]:=Sum[(-1)^k Binomial[j+n, k] Binomial[j-n,j+m-k], {k,0,j+m}]
MatrixForm[Table[Sum[(-I)^k K[m,k] K[k,n], {k,-j,j}] - (-2)^j I^(m+n) K[m,n], {m,-j,j}, {n,-j,j}]]
\end{verbatim}
}
\noindent for $j\!\in \! \left\{ 0,1,2,3,...\right\}$, and the program
{\small
\begin{verbatim}
j=3/2   
K[m_,n_]:=Sum[(-1)^k Binomial[j+n, k] Binomial[j-n,j+m-k], {k,0,j+m}]
N[MatrixForm[Table[Sum[(-I)^k K[m,k] K[k,n], {k,-j,j}] - (-2)^j I^(m+n) K[m,n], {m,-j,j}, {n,-j,j}]]]
\end{verbatim}
}
\noindent for $j\!\in \! \left\{\frac{1}{2},\frac{3}{2},\frac{5}{2},...\right\}$.\\[5mm]

\item
\noindent By using the program
{\small
\begin{verbatim}
j = 10
KP[m_, n_] :=  Sum[(-1)^k Binomial[j + n, k] Binomial[j - n, j + m - k], {k, 0,  j + m}]
KF[m_, n_] := (1/2^j) Sqrt[Binomial[2 j, j + n]/Binomial[2 j, j - m]] KP[m, n]
K := Table[(-1)^(2 j) I^m  KF[-m, n], {m, -j, j}, {n, -j, j}];
Jz := Table[m DiscreteDelta[m - n], {m, -j, j}, {n, -j, j}]
Jx := K.K.Jz.K
Jy := K.Jz.K.K
MatrixForm[MatrixPower[K, 3]]
MatrixForm[Jz]
MatrixForm[Jx]
MatrixForm[Jy]
MatrixForm[Jx.Jy - Jy.Jx - I Jz]
MatrixForm[Jy.Jz - Jz.Jy - I Jx]
MatrixForm[Jz.Jx - Jx.Jz - I Jy]
\end{verbatim}
}
\noindent for $j\!\in \! \left\{ 0,1,2,3,...\right\}$, and
{\small
\begin{verbatim}
j = 3/2
KP[m_, n_] :=  Sum[(-1)^k Binomial[j + n, k] Binomial[j - n, j + m - k], {k, 0,  j + m}]
KF[m_, n_] := (1/2^j) Sqrt[Binomial[2 j, j + n]/Binomial[2 j, j - m]] KP[m, n]
K := Table[(-1)^(2 j) I^m  KF[-m, n], {m, -j, j}, {n, -j, j}];
Jz := Table[m DiscreteDelta[m - n], {m, -j, j}, {n, -j, j}]
Jx := K.K.Jz.K
Jy := K.Jz.K.K
N[MatrixForm[MatrixPower[K, 3]]]
N[MatrixForm[Jz]]
N[MatrixForm[Jx]]
N[MatrixForm[Jy]]
N[MatrixForm[Jx.Jy - Jy.Jx - I Jz]]
N[MatrixForm[Jy.Jz - Jz.Jy - I Jx]]
N[MatrixForm[Jz.Jx - Jx.Jz - I Jy]]
\end{verbatim}
}
\noindent \noindent for $j\!\in \! \left\{\frac{1}{2},\frac{3}{2},\frac{5}{2},...\right\}$, one can directly verify the relation
\[
K^3=\mathbb{I}
\]
and to see that the operators
\[
J_x=K^+QK, \qquad J_y=KQK^+\quad \mbox{and}\quad J_z=Q
\]
satisfy the relations
\[
[J_x,J_y]={\rm i}J_z,\quad [J_y,J_z]={\rm i}J_x,\quad [J_z,J_x]={\rm i}J_y.
\]
\item
\noindent
It is known that some identities for powers fail for complex numbers, no matter how complex powers are defined as single-valued functions. For example:
\[  \begin{array}{l}
\begin{array}{r}
(-1)^{\frac{1}{2}}={\rm e}^{\frac{\pi {\rm i}}{2}}\\
{\rm i}^{\frac{1}{2}}={\rm e}^{\frac{\pi {\rm i}}{4}}\\
(-{\rm i})^{\frac{1}{2}}={\rm e}^{-\frac{\pi {\rm i}}{4}}
\end{array}\qquad \text{and consequently}\qquad 
(-1)^{\frac{1}{2}}\ {\rm i}^{\frac{1}{2}}\neq (-{\rm i})^{\frac{1}{2}};\\[8mm]
\begin{array}{r}
(-1)^{\frac{1}{2}}={\rm e}^{\frac{\pi {\rm i}}{2}}={\rm i}\\
1^{\frac{1}{2}}=1
\end{array}\qquad \text{and consequently}\qquad 
\left(\frac{1}{-1}\right)^{\frac{1}{2}}\neq \frac{1^{\frac{1}{2}}} {(-1)^{\frac{1}{2}}}.
\end{array}
\]
So, we have to be careful in computations involving half-integer powers of complex non-positive numbers. \\
In our computations involving half-integer powers of \ $-1$, \ ${\rm i}$, \ $-{\rm i}$, \ we use:
\begin{itemize}
\item the definition \ $z^k=|z|^k \, {\rm e}^{{\rm i}k\, {\rm arg}(z)}$;
\item the identity  \ $z^k\ z^m\!=\!z^{k+m}$ \  for integer as well as half-integer $k$ and $m$;
\item the identities  \ $z_1^k\ z_2^k\!=\!(z_1\, z_2)^{k}$ \ and \ $\frac{z_1^k} {z_2^k}\!=\!\left(\frac{z_1}{z_2}\right)^{k}$ \  only for integer  $k$.
\end{itemize}

\item
The proofs of our theorems are based on the relations:
\[
\begin{array}{l}
(1-{\rm i})^{2j}=(\sqrt{2})^{2j}\, {\rm e}^{-\frac{2j\pi {\rm i}}{4}}=2^j\,  {\rm e}^{-\frac{j\pi {\rm i}}{2}}=2^j\, (-{\rm i})^j;\\[5mm]
\begin{array}{rl}
(1+{\rm i})^{j+n}\, (1-{\rm i})^{j-n} 
 & \!\!\!=\left(\frac{1+{\rm i}}{1-{\rm i}}\right)^{j+n}\, (1-{\rm i})^{2j}
 ={\rm i}^{j+n}(1-{\rm i})^{2j}=2^j\, (-{\rm i})^j\, {\rm i}^{j+n};
\end{array}\\[5mm]
\begin{array}{rl}
{\rm i}^{m+k}\, (-1)^{j+k}\, (-1)^{j+n} & \!\!\!=(-1)^{j+n}\, {\rm i}^{m-j} \, {\rm i}^{j+k}\, (-1)^{j+k}\\[2mm]
 & \!\!\!=(-1)^{j+n}\, {\rm i}^{m-j}\, (-{\rm i})^{j+k} 
 =(-1)^{j+n}\, {\rm i}^{m-j}\, (-{\rm i})^{j}\,(-{\rm i})^{k};
\end{array}\\[5mm]
\begin{array}{rl}
(-1)^{j+n}\, {\rm i}^{m-j}\, (-{\rm i})^{j}\, {\rm i}^{j+m}\, {\rm i}^{j+n} & \!\!\!=(-1)^{j+n}\, {\rm i}^{2m}\, {\rm i}^{j+n}\, (-{\rm i})^{j}=(-1)^{j+m+n}\, {\rm i}^{j+n}\, (-{\rm i})^{j}\\[2mm]
 & \!\!\!=(-1)^n\, {\rm i}^{j+n}\, (-{\rm i})^{j}\, (-1)^{j+m}
\end{array}\\[5mm]
\begin{array}{rl}
(-1)^n\, {\rm i}^{j+n}\, (-{\rm i})^{j} & \!\!\!=
\frac{1}{(-1)^j}\, (-1)^{j+n}\, {\rm i}^{j+n}\, (-{\rm i})^{j} 
=\frac{1}{(-1)^j}\, (-{\rm i})^{j+n}\, (-{\rm i})^{j} 
=\frac{1}{(-1)^j}\, (-{\rm i})^{2j}\, (-{\rm i})^{n} \\[2mm]
 & \!\!\!=\frac{(-1)^j}{(-1)^{2j}}\, (-{\rm i})^{2j}\, (-{\rm i})^{n}
=(-1)^j\, \frac{(-{\rm i})^{2j}}{(-1)^{2j}} \, (-{\rm i})^{n}
=(-1)^j\, \left(\frac{-{\rm i}}{-1}\right)^{2j} \, (-{\rm i})^{n}
=(-1)^{2j}\,  (-{\rm i})^{n}.
\end{array}

\end{array}
\]

\end{itemize}

\end{widetext}
\end{document}